\documentclass[conference,10pt]{IEEEtran}
\usepackage{amsmath,amssymb,amsfonts}
\usepackage{algorithm}
\usepackage{array}
\usepackage{textcomp}
\usepackage{stfloats}
\usepackage{url}
\usepackage{verbatim}
\usepackage{graphicx}

\usepackage{tabularx}

\usepackage[style=ieee, backend=bibtex]{biblatex}
\usepackage{algpseudocode}
\usepackage{subcaption}
\usepackage{hhline}
\usepackage{multicol}
\usepackage{multirow}
\usepackage[T1]{fontenc}
\usepackage{orcidlink}

\usepackage{threeparttable}

\usepackage[utf8]{inputenc}
\usepackage{pgfplots}
\DeclareUnicodeCharacter{2212}{−}
\usepgfplotslibrary{groupplots,dateplot}
\usetikzlibrary{patterns,shapes.arrows}
\pgfplotsset{compat=newest}
\pgfplotsset{compat=1.11,
    /pgfplots/ybar legend/.style={
    /pgfplots/legend image code/.code={%
       \draw[##1,/tikz/.cd,yshift=-0.25em]
        (0cm,0cm) rectangle (15pt,0.8em);},
   },
}

\begin{document}

\title{GTQCP: Greedy Topology-Aware Quantum Circuit Partitioning}

\author{\IEEEauthorblockN{Joseph Clark \orcidlink{0009-0000-0295-0083}}
\IEEEauthorblockA{\textit{Dept. of Elec. Eng. and Computer Science} \\
\textit{University of Tennessee}\\
Knoxville, USA \\
jclar168@vols.utk.edu}
\and
\IEEEauthorblockN{Travis S. Humble \orcidlink{0000-0002-9449-0498}}
\IEEEauthorblockA{\textit{Quantum Science Center} \\
\textit{Oak Ridge National Laboratory} \\
Oak Ridge, USA \\
humblets@ornl.gov}
\and
\IEEEauthorblockN{Himanshu Thapliyal \orcidlink{0000-0001-9157-4517}}
\IEEEauthorblockA{\textit{Dept. of Elec. Eng. and Computer Science} \\
\textit{University of Tennessee}\\
Knoxville, USA \\
hthapliyal@utk.edu}
}

\maketitle

\begin{abstract}
We propose Greedy Topology-Aware Quantum Circuit Partitioning (GTQCP), a novel quantum gate circuit partitioning method which partitions circuits by applying a greedy heuristic to the qubit dependency graph of the circuit. GTQCP is compared against three other gate partitioning methods, two of which (QuickPartitioner and ScanPartitioner) are part of the Berkley Quantum Synthesis Toolkit. GTQCP is shown to have 18\% run time improvement ratio over the fastest approach (QuickPartitioner), and a 96\% improvement over the highest quality approach (ScanPartitioner). The algorithm also demonstrates nearly identical result quality (number of partitions) compared with ScanPartitioner, and a 38\% quality improvement over QuickPartitioner.
\end{abstract}


\section{Introduction}
\IEEEPARstart{Q}{uantum} computing is an emerging computing paradigm which promises to provide advancements in medicine, physics, and mathematics. The theoretical advantage of these machines is the result of leveraging the properties of quantum mechanics, which allows quantum algorithms to solve problems faster than existing classical techniques. However, the design and implementation of quantum circuits large enough to produce a quantum advantage has proven difficult due to the limitations of Noisy Intermediate-Scale Quantum (NISQ) computers and the computational complexity associated with manipulating quantum circuits.

Quantum circuit partitioning presents a method for circumventing some of the complexity associated with quantum computing. In quantum circuit partitioning, a circuit is divided into sub-circuits which are either easier to execute on quantum hardware \cite{cutqc, dynamicpartitioning, automatedhypergraphpartitioning, kl_partitioning} or easier to manipulate using classical computers \cite{QGo, QEst, quickish_partitioning}. Quantum circuit partitioning algorithms are divided into two broad categories, depending on whether they focus on partitioning qubits or gates. Qubit partitioning methods \cite{evolutionary_partitioning, kl_partitioning} assign qubits to partitions such that the number of non-local gate operations is minimized. Gate partitioning methods assign gates to partitions, such that the number of teleportation operations is minimized. Because qubit partitioning "splits" some multi-qubit gates when creating partitions, circuits partitioned this way remain difficult to process with classical methods. Gate partitioning \cite{QGo, QEst, kl_partitioning, quickish_partitioning}, on the other hand, does not split multi-qubit gates, and thus may be used for both quantum and classical applications.

An efficient method for partitioning quantum circuits is balanced min-cut, which can be modified to perform either qubit or gate partitioning \cite{kl_partitioning}. However, as a global partitioning method, balanced min-cut does not consider local qubit connectivity patterns, which reduces result quality \cite{teleport_explanation}. Thus, a local method, or a global method augmented with localized information, is preferable for generating an optimal result.

We choose to focus on quantum circuit partitioning for peephole optimization, which we define as the gate partitioning of an $n$-qubit circuit with $g$ gates into the \underline{minimum number of sub-circuits}, $j$, such that each sub-circuit contains $\leq k$ qubits. Minimizing the number of partitions maximizes the amount of information available in each peephole during optimization, which improves result quality. Limiting the number of qubits in each sub-circuit reduces the complexity of conventional quantum circuit synthesis from $O(4^n)$ to $O(4^k)$, which allows the method to be applied to arbitrarily large circuits. There are several existing approaches to this problem \cite{quickish_partitioning, merged_kl}, in this work we have compared against those offered in Lawrence Berkeley National Laboratory's Berkeley Quantum Synthesis Toolkit (BQSKit) \cite{BQSKitUpdated}.

BQSKit provides two notable methods of performing circuit partitioning for peephole optimization. The first is ScanPartitioner, which was proposed as the partitioning method for QGo \cite{QGo}. ScanPartitioner applies a greedy approach combined with exhaustive search to iteratively create partitions. This approach produces high quality results, but has a very high time complexity of $O(gn^k)$. The second method from BQSKit is QuickPartitioner \cite{BQSKitUpdated}, which iterates over gates in topological order and sorts them into partitions. QuickPartitioner has a significantly better time complexity of approximately $O(gn)$, but with signficantly lower quality than ScanPartitioner. A third method is our prior work \cite{PreviousPaper}, which uses a tree-based approach which operates on the Directed Acyclic Graph (DAG) representation of the circuit to find only groups of qubits which interact in the local region of the circuit. With a time complexity of $O(gn4^k)$, this method demonstrates nearly identical performance to ScanPartitioner while showing a considerable speedup for large circuits or high values of $k$. However, this approach still scales poorly with increasing $k$, and fails to find optimal solutions on shallow circuits.

We propose a novel partitioning method for peephole optimization called Greedy Topology-Aware Quantum Circuit Partitioning (GTQCP), which uses a greedy heuristic based on the qubit dependency graph of the circuit to generate partitions. The algorithm takes as input an $n$ qubit circuit with $g$ gates and a parameter $k \leq n$, and outputs a partitioned version of the circuit with the minimal number of partitions such that each partition contains no more than $k$ qubits. GTQCP produces results of nearly identical quality to ScanPartitioner and \cite{PreviousPaper}, and yet has a time complexity of no more than  $O(gne^\frac{k}{e})$ compared with ScanPartitioner's $O(gn^k)$ and our prior work's $O(gn4^k)$. GTQCP outperforms QuickPartitioner in quality by 38\%, and even in run time by 18\%. The run time advantage of GTQCP over ScanPartitioner and \cite{PreviousPaper} is even larger at 96\% and 70\%, respectively. In a more detailed analysis of the behavior of the four methods on several test circuits for larger values of $k$, GTQCP demonstrates a sub-exponential run time growth rate for all circuits, unlike ScanPartitioner and \cite{PreviousPaper}. Also, for some circuits, the quality advantage for the higher quality methods (including GTQCP) against QuickParitioner is as high as 69\%. GTQCP also addresses the shortcoming in our prior work which limits result quality on shallow circuits.

This paper is organized as follows: Section \ref{prop_work} describes the behavior of GTQCP. Section \ref{res} discusses how the partitioners were evaluated and the results. Section \ref{conc} discusses future research direction and concludes the article.

\section{Proposed Method: Greedy Topology-Aware Quantum Circuit Partitioning (GTQCP)}
\label{prop_work}

GTQCP partitions a circuit with $n$ qubits and $g$ gates into the minimum number of subcircuits with no more than $k$ qubits. The algorithm follows three steps which are broadly similar to ScanPartitioner and our prior work. 1) First, the algorithm produces a set of candidate groups of at most $k$ qubits. 2) Next, these groups are expanded into candidate partitions by accumulating gates which depend only on the target qubits. 3) The candidate partitions are then scored, and the candidate with the best score is reserved as a partition and removed from consideration for future partitions. This process is repeated until all gates have grouped into a partition. The latter two steps are very similar between GTQCP, ScanPartitioner, and \cite{PreviousPaper}, with step 2 consisting of simple iteration from the start of the circuit for each group, and step 3 scoring partitions by the number of circuit gates. Step 1 is implemented significantly differently by all three methods, with a substantial effect on runtime and time complexity. ScanPartitioner produces all possible groups of qubits by calculating all simple paths through the qubit coupling graph of the circuit, which has a time complexity of $O(gn^k)$. Our previous work \cite{PreviousPaper} improves on this design by calculating only the qubit groups capable of interacting in the graph representation of the circuit on each iteration, producing a time complexity of $O(gn4^k)$. GTQCP improves on this design further by greedily collecting qubits on the dependency graph of the circuit, producing a time complexity of $O(gne^\frac{k}{e})$.

\begin{figure}[t]
\centering
\includegraphics[trim= 0cm 0cm 0cm 0cm, width=\columnwidth]{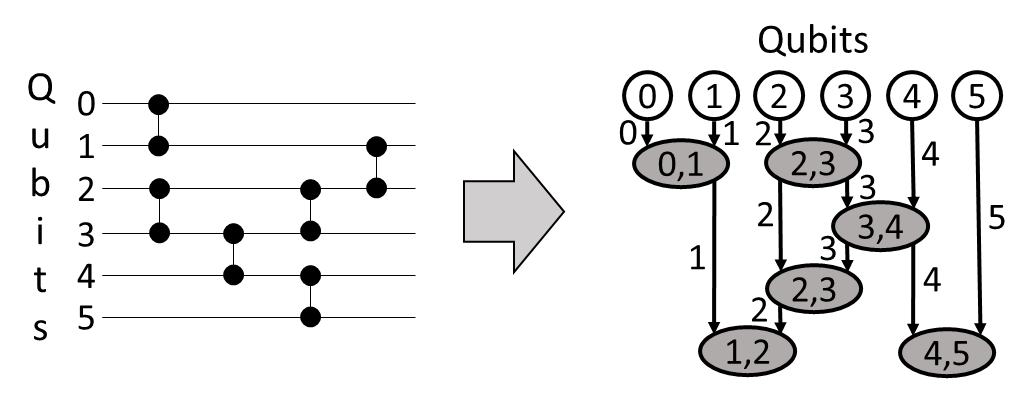}
\caption{Converting a Quantum Circuit into a Direct Acyclic Graph}
\label{DAG_conversion}
\end{figure}

\subsection{Gate Dependencies}

\begin{figure}[t]
\centering
\includegraphics[trim= 0cm 0cm 0cm 0cm, width=0.6\columnwidth]{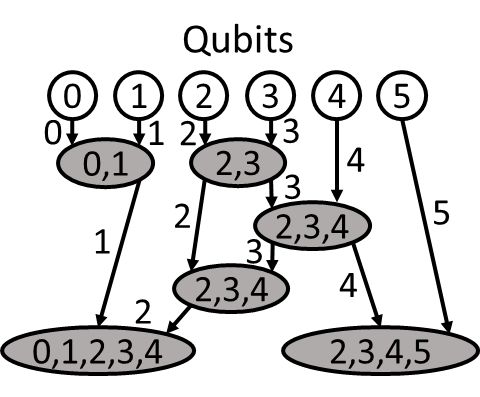}
\caption{Gate Dependency Calculation}
\label{dependency_calculation}
\end{figure}

Both our prior work and GTQCP use a qubit dependency graph of the circuit as part of their partitioning operation. The dependency graph is produced by performing a modified breadth-first search over the DAG representation of the circuit. All qubit nodes are considered as a starting point, after which the nodes in the queue are visited in topological order. The dependencies of a node are merged into each of its children when the node is visited, which propagates dependencies forward. The result of applying the algorithm to the example circuit is shown in Figure \ref{dependency_calculation}. The first step visits each qubit node and merges the qubit into the dependencies for the first gate on each qubit. Each visited gate is added to the search queue.

In the first step, each qubit node is visited and merged into the list of dependencies for the first gate along the qubit. Gates are added to the visitation queue when they are encountered.
The first gates visited are the gate between qubits 0 and 1 and the gate between qubits 2 and 3, which have dependencies $\{0,1\}$ and $\{2,3\}$, respectively. These dependencies are copied to their direct children to form: $\{0,1,2\}$ for the gate between 1 and 2, $\{2,3\}$ for the second gate between qubits  2 and 3, and $\{2,3,4\}$ for the gate between qubits 3 and 4. These gates are visited next and their dependencies are similarly copied into each of their children. After another round, the algorithm terminates because all remaining gates are dependent on at least $k$ qubits.

\subsection{Qubit Group Calculation}

\begin{figure*}
\centering
\begin{subfigure}{0.32\textwidth}
	\centering
	\includegraphics[trim= 0cm 0cm 0cm 0cm, width=\dimexpr\textwidth-0.05\textwidth\relax]{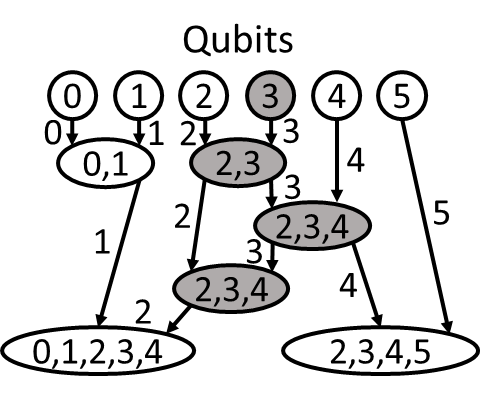}
	\caption{Step 1: Longest Path for Qubit 3}
\end{subfigure}
\vrule
\begin{subfigure}{0.32\textwidth}
	\centering
	\includegraphics[trim= 0cm 0cm 0cm 0cm, width=\dimexpr\textwidth-0.05\textwidth\relax]{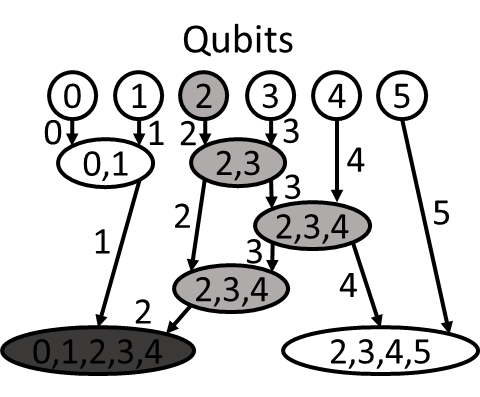}
	\caption{Step 2: Longest Path for Qubit 2}
\end{subfigure}
\vrule
\begin{subfigure}{0.32\textwidth}
	\centering
	\includegraphics[trim= 0cm 0cm 0cm 0cm, width=\dimexpr\textwidth-0.05\textwidth\relax]{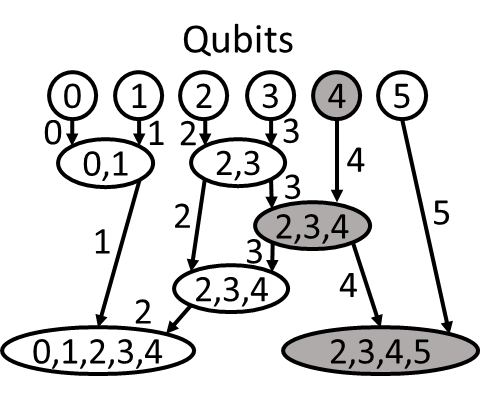}
	\caption{Step 3: Longest Path for Qubit 4}
\end{subfigure}
\caption{Group Enumeration Steps on Example Circuit for Qubit 3}
\label{path_enumeration}
\end{figure*}

The qubit group calculation employed by GTQCP improves on our prior work by using an improved heuristic which produces a smaller worst-case number of groups. Our prior work identified qubit groups by considering all possible paths through the DAG of a given circuit; the improved method considers only the longest path available along each qubit such that the group size does not exceed $k$. Figure \ref{path_enumeration} shows the behavior of the algorithm on the example circuit. We choose to start by finding the longest path down qubit 3 (step 1). The dependencies for the longest path down qubit 3 are $\{2,3,4\}$, with new qubits $\{2,4\}$. Next, the longest path down qubit 2 is found (step 2). Since $k=4$ for our example, the gate in the bottom left (shaded dark gray) is not included, and qubit 2 yields $\{2,3,4\}$. As this contains another unexplored qubit (4), that qubit is explored (step 3), yielding $\{2,3,4,5\}$. This is a full group, so the algorithm move backwards until there is another qubit to explore. This returns to the group at the end of qubit 3, where 4 is explored, finding $\{2,3,4,5\}$ again. As there are no more unexplored qubits, the enumeration along qubit 3 is complete.

\begin{algorithm}
\caption{Qubit Group Enumeration for GTQCP}
\label{group_enumeration}
\begin{algorithmic}[1]
\Function{EnumerateGroups}{$circuit, depend$}
	\State $results \gets \emptyset$
	\ForAll{$qubit \in circuit$}
		\State \Call{Enumerate}{$qubit, set\{qubit\}$}
	\EndFor
	\State \textbf{return} $results$
\EndFunction
\State
\Function{Enumerate}{$target, input$}
	\State $gate \gets target$
	\While{$|input \cup depend[next(gate)]| \leq k$}
		\State $gate \gets next(gate)$
	\EndWhile
	\State $group \gets input \cup depend[gate]$
	\If{$group \notin results$}
		\State $results.add(group)$
		\If{$|group| < k$}
			\ForAll{$qubit \in (group - input)$}
				\State \Call{Enumerate}{$qudit, input \cup set\{qubit\}$}
			\EndFor
		\EndIf
	\EndIf
	
\EndFunction
\end{algorithmic}
\end{algorithm}

Algorithm \ref{group_enumeration} shows the qubit group calculation algorithm. The algorithm accepts the circuit to be partitioned and the gate dependencies as inputs, and produces a list of candidate groups as output. Groups are generated by greedily finding the gate along a qubit which depends on the largest number of qubits not greater than $k$ (line 11 to 13). The resulting group is added to the results (line 16), and each new qubit in the group is recursively explored the same way (line 18 to 20). Additionally, a small optimization is applied which prevents the algorithm from recurring on already explored groups by checking if the group is already in the result set (line 15). This process is repeated for each qubit in the circuit (line 3 to 5), and the result set is returned when the process is finished.

\subsection{Complexity}

GTQCP improves on the time complexity of both ScanPartitioner and our previous work by employing a greedy strategy which does not affect the overall performance of the algorithm. We can model this process as a graph, wherein each node represents the grouping at the end of a qubit. Each node has $b$ qubits out of the $k$ qubits allowed, such that the graph is $d$ nodes deep. Given these values, we can calculate that there will be $b^d = t$ possible groupings. Because the qubits must be spread across the depth of the graph, $\frac{k}{d} = b$, the total number of groupings is given by $b^\frac{k}{b} = t$. Optimizing the value of $b$ to produce the maximum number of groups produces $b = e$. Thus, $e^\frac{k}{e} = t$. This worst-case upper bound on time complexity in $k$ is better than both our prior work ($4^k$) and ScanPartitioner ($n^k$). This operation is repeated for each qubit ($n$ times) and will, at worst, visit each gate on each iteration. Thus, the overall worst-case upper bound on complexity is $O(gne^\frac{k}{e})$.

It should be noted that the this worst-case upper bound is very pessimistic and does not occur in any of the benchmark tests. In fact, due to the optimizations applied in this algorithm, the time complexity for all tested circuit structures does not appear to be exponential. Instead, our tests show constant or linear response to increasing $k$, as demonstrated by the run time measurements. This is because the complexity of the algorithm is bounded in two directions: too little connectivity results in few branches on the search tree, and too much causes the group to fill up quickly and produce a shallow tree. Additionally, although the reduction in the worst case number of groups necessarily means that some qubit groups detected by our prior work will not be found by this approach, the quality of result for this method, as demonstrated in the results section, does not appear to be impaired by this limitation.

\section{Results}
\label{res}

GTQCP, our partitioner in \cite{PreviousPaper}, ScanPartitioner, and QuickPartitioner were applied to a set of benchmark circuits and the run time and the number of partitions produced were measured for each run. The CNOT gate count, qubit count, and a brief description of each circuit is provided in Table \ref{circuit_info}. The analysis was performed at $k = 4$ and $5$, because 4 and 5 are common values of this parameter for this application. All tests were performed using a computer with an AMD Ryzen 5 5600X processor and 32GB of RAM.

Because \cite{PreviousPaper} and GTQCP do not group together qubits which do not interact in the active part of the circuit, a simple reprocessing algorithm is applied to the results for each method which combines adjacent blocks containing a total of no more than $k$ qubits. This is done in order to provide an accurate comparison between the four methods.

\begin{table*}[th!]
\caption{Benchmark Circuits Used in This Work}
\begin{tabularx}{\textwidth}{|X|l|X|X|}
\hline
             Circuit &    Description &    CNOT Count &   Qubit Count \\
\hline
             adder\_9 & Quantum adder                                    &   98 &   9 \\
        heisenberg\_8 & 50 step Heisenberg model simulation              & 2100 &   8 \\
              hlf\_10 & Hidden linear function circuit                   &   56 &  10 \\
         multiply\_10 & Quantum multiplier                               &  163 &  10 \\
             qaoa\_10 & Quantum approximate optimization algorithm       &   85 &  10 \\
               qft\_5 & Quantum Fourier transform circuit                &   33 &   5 \\
              qft\_10 & Quantum Fourier transform circuit                &  216 &  10 \\
              qft\_20 & Quantum Fourier transform circuit                &  380 &  20 \\
              TFIM\_4 & 100 step transverse-field Ising model simulation &   12 &   4 \\
              TFIM\_8 & 100 step transverse-field Ising model simulation &   56 &   8 \\
             TFIM\_16 & 100 step transverse-field Ising model simulation &  240 &  16 \\
             TFIM\_32 & 100 step transverse-field Ising model simulation &  992 &  32 \\
           wstate\_27 & W-state preparation circuit                      &   52 &  27 \\
\hline
\end{tabularx}
\label{circuit_info}
\end{table*}

\begin{table*}
\centering
\caption{Benchmarks of Partitioning Methods for $k$ at $4$ and $5$}
\begin{tabularx}{\textwidth}{|l|>{\hsize=.5\hsize}X|XX|XX|XX|XX|}
\hline
 &  & \multicolumn{2}{>{\hsize=2\hsize}X|}{Quick \cite{BQSKitUpdated}} & \multicolumn{2}{>{\hsize=2\hsize}X|}{Scan \cite{QGo}} & \multicolumn{2}{>{\hsize=2\hsize}X|}{Clark et al. \cite{PreviousPaper}} & \multicolumn{2}{>{\hsize=2\hsize}X|}{GTQCP (Proposed)} \\
\hhline{|~|~|*8{-}}
           Circuit &    k & Time (s) & Partitions & Time (s) & Partitions & Time (s) & Partitions & Time (s) &  Partitions \\
\hline
     adder\_9 & 4 &                       0.04 &                         15 &            0.04 &               7 &                              0.07 &                                 7 &  0.04 &     7 \\
     adder\_9 & 5 &                       0.03 &                          7 &            0.06 &               6 &                              0.08 &                                 6 &  0.03 &     6 \\
heisenberg\_8 & 4 &                       1.18 &                        349 &            0.99 &             225 &                              2.19 &                               225 &  0.99 &   225 \\
heisenberg\_8 & 5 &                       1.24 &                        294 &            1.00 &             150 &                              2.75 &                               150 &  0.98 &   150 \\
      hlf\_10 & 4 &                       0.02 &                         15 &            0.03 &               8 &                              0.04 &                                 8 &  0.02 &     8 \\
      hlf\_10 & 5 &                       0.02 &                         10 &            0.05 &               5 &                              0.04 &                                 5 &  0.02 &     5 \\
 multiply\_10 & 4 &                       0.06 &                         19 &            0.10 &              15 &                              0.13 &                                15 &  0.05 &    15 \\
 multiply\_10 & 5 &                       0.05 &                         11 &            0.12 &               8 &                              0.14 &                                 8 &  0.05 &     8 \\
     qaoa\_10 & 4 &                       0.03 &                         16 &            0.05 &               9 &                              0.06 &                                 9 &  0.02 &     9 \\
     qaoa\_10 & 5 &                       0.03 &                          9 &            0.08 &               6 &                              0.07 &                                 6 &  0.02 &     6 \\
       qft\_5 & 4 &                       0.01 &                          3 &            0.01 &               3 &                              0.02 &                                 3 &  0.01 &     3 \\
       qft\_5 & 5 &                       0.01 &                          1 &            0.01 &               1 &                              0.02 &                                 1 &  0.01 &     1 \\
      qft\_10 & 4 &                       0.08 &                         27 &            0.12 &              18 &                              0.17 &                                18 &  0.07 &    18 \\
      qft\_10 & 5 &                       0.07 &                         17 &            0.16 &              12 &                              0.20 &                                12 &  0.07 &    12 \\
      qft\_20 & 4 &                       0.21 &                         71 &           30.82 &              45 &                              0.52 &                                45 &  0.16 &    45 \\
      qft\_20 & 5 &                       0.21 &                         51 &          152.29 &              33 &                              0.63 &                                32 &  0.17 &    35 \\
      TFIM\_8 & 4 &                       0.05 &                          8 &            0.05 &               7 &                              0.09 &                                 7 &  0.04 &     7 \\
      TFIM\_8 & 5 &                       0.05 &                          5 &            0.05 &               5 &                              0.10 &                                 5 &  0.05 &     5 \\
     TFIM\_16 & 4 &                       0.22 &                         44 &            0.22 &              31 &                              0.51 &                                32 &  0.18 &    30 \\
     TFIM\_16 & 5 &                       0.25 &                         35 &            0.23 &              22 &                              0.58 &                                22 &  0.18 &    22 \\
     TFIM\_32 & 4 &                       0.96 &                        184 &            0.90 &             125 &                              3.49 &                               126 &  0.82 &   124 \\
     TFIM\_32 & 5 &                       1.00 &                        134 &            0.92 &              86 &                              3.80 &                                87 &  0.80 &    84 \\
   wstate\_27 & 4 &                       0.03 &                         19 &            0.04 &              13 &                              0.11 &                                26 &  0.03 &    13 \\
   wstate\_27 & 5 &                       0.03 &                         14 &            0.04 &               9 &                              0.10 &                                18 &  0.03 &     9 \\
\hline
\end{tabularx}
\label{results_with_small_k}
\end{table*}

\begin{table*}
\centering
\caption{Performance Improvement of GTQCP Compared with Existing Works}
\begin{tabularx}{\textwidth}{|l|>{\hsize=.5\hsize}c|X|XX|}
\hline
 &   & Partitions & \multicolumn{2}{>{\hsize=2\hsize}c|}{Time (s)} \\
\hhline{|~|~|*3{-}}
 & k & GTQCP w.r.t. Quick \cite{BQSKitUpdated} & GTQCP w.r.t. Scan \cite{QGo} & GTQCP w.r.t. Clark et al. \cite{PreviousPaper} \\
\hline
     Improvement Ratio &    4 &             34.55\% &            92.72\% &             67.22\% \\
     Improvement Ratio &    5 &             41.67\% &            98.45\% &             71.80\% \\
\hline
\multicolumn{2}{|>{\hsize=2\hsize}c|}{\textbf{Average Improvement Ratio}} & \textbf{38.11\%} & \textbf{95.58\%} & \textbf{69.51\%} \\
\hline
\end{tabularx}
\label{performance_with_small_k}
\end{table*}

Table \ref{results_with_small_k} shows the performance data for the partitioners for all benchmark circuits, while Table \ref{performance_with_small_k} presents a summary of the data. The results show that GTQCP produces results of equal quality to ScanPartitioner and \cite{PreviousPaper}, with all three showing a 38\% quality improvement against QuickPartitioner. GTQCP is faster than all three other partitioners for most test circuits, with an average run time improvement of 18\%, 96\%, and 70\% against QuickPartitioner, ScanPartitioner, and \cite{PreviousPaper}, respectively. Interestingly, although the loose upper bound for GTQCP is significantly higher than QuickPartitioner, similar run times and growth rates are observed for both methods across the tested circuits.

\subsection{Performance on Larger Values of $k$}

A more detailed analysis was also performed on four of the more important benchmark circuits (multiply\_10, qaoa\_10, qft\_20, and TFIM\_32) for values of k from $3$ to $16$. Once again, the average execution time and number of partitions produced were measured for each benchmark.

The results are shown in Figure \ref{detailed_performance}. The range of $k$ values and circuit structures tested gives an indication of the general performance of the four algorithms. For example, although ScanPartitioner is tied as the best quality method in all cases, it also demonstrates exponential growth in run time for every circuit. Similarly, although QuickParitioner is one of the fastest methods in all cases, it also produces the worst quality results in every case, producing comparable results to the other methods only on the multiply circuit. In the TFIM circuit, which is the most extreme case, QuickPartitioner produces as many as three times the number of partitions that the other three methods generate. Our partitioner in \cite{PreviousPaper} produces similar performance to GTQCP for most circuits, with good result quality and only slightly higher run time. However, \cite{PreviousPaper} does demonstrate exponential growth on the TFIM circuit, while GTQCP appears linear. 

\begin{figure*}
\centering
\begin{subfigure}[t]{0.49\textwidth}
	\centering
	\pgfplotsset{every axis legend/.append style={font=\fontsize{9}{8.25}\selectfont}}
	\def\axisdefaultwidth{0.8\textwidth}
	\def\axisdefaultheight{0.55\textwidth}
\begin{tikzpicture}

\definecolor{crimson2143940}{RGB}{214,39,40}
\definecolor{darkgray176}{RGB}{176,176,176}
\definecolor{darkturquoise23190207}{RGB}{23,190,207}
\definecolor{forestgreen4416044}{RGB}{44,160,44}
\definecolor{lightgray204}{RGB}{204,204,204}
\definecolor{mediumpurple148103189}{RGB}{148,103,189}

\begin{groupplot}[group style={group size=1 by 2, x descriptions at=edge bottom, vertical sep = 0pt}]
\nextgroupplot[
log basis y={10},
tick align=outside,
tick pos=left,
title={multiply\_10},
x grid style={darkgray176},
xmin=2.65, xmax=10.35,
xtick style={color=black},
y grid style={darkgray176},
ylabel={Time (ms)},
ymin=37.1558842163936, ymax=2972.63143197904,
ymode=log,
ytick style={color=black},
ytick={1,0.2,0.5,10,2,5,100,20,50,1000,200,500,10000,2000,5000,100000,20000,50000},
yticklabels={
  1\(\displaystyle \cdot10^{0}\),
  2\(\displaystyle \cdot10^{-1}\),
  5\(\displaystyle \cdot10^{-1}\),
  1\(\displaystyle \cdot10^{1}\),
  2\(\displaystyle \cdot10^{0}\),
  5\(\displaystyle \cdot10^{0}\),
  1\(\displaystyle \cdot10^{2}\),
  2\(\displaystyle \cdot10^{1}\),
  5\(\displaystyle \cdot10^{1}\),
  1\(\displaystyle \cdot10^{3}\),
  2\(\displaystyle \cdot10^{2}\),
  5\(\displaystyle \cdot10^{2}\),
  1\(\displaystyle \cdot10^{4}\),
  2\(\displaystyle \cdot10^{3}\),
  5\(\displaystyle \cdot10^{3}\),
  1\(\displaystyle \cdot10^{5}\),
  2\(\displaystyle \cdot10^{4}\),
  5\(\displaystyle \cdot10^{4}\)
}
]
\addplot [semithick, forestgreen4416044, mark=x, mark size=3, mark options={solid}]
table {%
3 61.78368
4 55.7779
5 52.13115
6 68.29612
7 51.52115
8 56.36903
9 55.4524
10 56.28264
};
\addplot [semithick, crimson2143940, mark=+, mark size=3, mark options={solid}]
table {%
3 100.85583
4 100.71311
5 123.27378
6 168.74454
7 291.73073
8 630.34287
9 1413.11868
10 2435.76793
};
\addplot [semithick, darkturquoise23190207, mark=*, mark size=3, mark options={solid}]
table {%
3 69.82543
4 64.12677
5 59.31354
6 60.33826
7 60.98627
8 77.80428
9 57.01965
10 58.52308
};
\addplot [semithick, mediumpurple148103189]
table {%
3 58.99128
4 47.75236
5 52.32406
6 46.47271
7 46.84133
8 54.24621
9 45.34535
10 45.99753
};

\nextgroupplot[
legend cell align={left},
legend style={fill opacity=0.8, draw opacity=1, text opacity=1, draw=lightgray204},
tick align=outside,
tick pos=left,
x grid style={darkgray176},
xlabel={k},
xmin=2.65, xmax=10.35,
xtick style={color=black},
y grid style={darkgray176},
ylabel={Partitions},
ymin=0, ymax=33.55,
ytick style={color=black}
]
\addplot [semithick, forestgreen4416044, opacity=0.7, mark=x, mark size=3, mark options={solid}]
table {%
3 32
4 19
5 11
6 8
7 5
8 5
9 3
10 1
};
\addlegendentry{QuickPartitioner \cite{BQSKitUpdated}}
\addplot [semithick, crimson2143940, opacity=0.7, mark=+, mark size=3, mark options={solid}]
table {%
3 29
4 15
5 8
6 6
7 5
8 4
9 2
10 1
};
\addlegendentry{ScanPartitioner \cite{BQSKitUpdated}}
\addplot [semithick, darkturquoise23190207, opacity=0.7, mark=*, mark size=3, mark options={solid}]
table {%
3 28
4 15
5 8
6 6
7 5
8 4
9 2
10 1
};
\addlegendentry{Clark et al. \cite{PreviousPaper}}
\addplot [semithick, mediumpurple148103189, opacity=0.7]
table {%
3 28
4 15
5 8
6 6
7 5
8 4
9 2
10 1
};
\addlegendentry{Proposed}
\end{groupplot}

\end{tikzpicture}
\end{subfigure}
\vrule
\begin{subfigure}[t]{0.49\textwidth}
	\centering
	\pgfplotsset{every axis legend/.append style={font=\fontsize{9}{8.25}\selectfont}}
	\def\axisdefaultwidth{0.8\textwidth}
	\def\axisdefaultheight{0.55\textwidth}
\begin{tikzpicture}

\definecolor{crimson2143940}{RGB}{214,39,40}
\definecolor{darkgray176}{RGB}{176,176,176}
\definecolor{darkturquoise23190207}{RGB}{23,190,207}
\definecolor{forestgreen4416044}{RGB}{44,160,44}
\definecolor{lightgray204}{RGB}{204,204,204}
\definecolor{mediumpurple148103189}{RGB}{148,103,189}

\begin{groupplot}[group style={group size=1 by 2, x descriptions at=edge bottom, vertical sep = 0pt}]
\nextgroupplot[
log basis y={10},
tick align=outside,
tick pos=left,
title={qaoa\_10},
x grid style={darkgray176},
xmin=2.65, xmax=10.35,
xtick style={color=black},
y grid style={darkgray176},
ylabel={Time (ms)},
ymin=18.0447570133592, ymax=2293.54185791386,
ymode=log,
ytick style={color=black},
ytick={1,0.2,0.5,10,2,5,100,20,50,1000,200,500,10000,2000,5000,100000,20000,50000},
yticklabels={
  1\(\displaystyle \cdot10^{0}\),
  2\(\displaystyle \cdot10^{-1}\),
  5\(\displaystyle \cdot10^{-1}\),
  1\(\displaystyle \cdot10^{1}\),
  2\(\displaystyle \cdot10^{0}\),
  5\(\displaystyle \cdot10^{0}\),
  1\(\displaystyle \cdot10^{2}\),
  2\(\displaystyle \cdot10^{1}\),
  5\(\displaystyle \cdot10^{1}\),
  1\(\displaystyle \cdot10^{3}\),
  2\(\displaystyle \cdot10^{2}\),
  5\(\displaystyle \cdot10^{2}\),
  1\(\displaystyle \cdot10^{4}\),
  2\(\displaystyle \cdot10^{3}\),
  5\(\displaystyle \cdot10^{3}\),
  1\(\displaystyle \cdot10^{5}\),
  2\(\displaystyle \cdot10^{4}\),
  5\(\displaystyle \cdot10^{4}\)
}
]
\addplot [semithick, forestgreen4416044, mark=x, mark size=3, mark options={solid}]
table {%
3 32.7167
4 30.04855
5 29.87633
6 27.6401
7 30.34323
8 30.39204
9 26.64844
10 25.73663
};
\addplot [semithick, crimson2143940, mark=+, mark size=3, mark options={solid}]
table {%
3 36.26995
4 47.20773
5 78.67049
6 101.90376
7 194.08438
8 474.12081
9 1086.39307
10 1840.1924
};
\addplot [semithick, darkturquoise23190207, mark=*, mark size=3, mark options={solid}]
table {%
3 34.21798
4 32.71862
5 34.24081
6 35.63645
7 32.1673
8 34.6124
9 33.8254
10 49.36733
};
\addplot [semithick, mediumpurple148103189]
table {%
3 26.25353
4 24.78399
5 24.07732
6 24.81187
7 23.09782
8 23.76407
9 22.61929
10 22.49026
};

\nextgroupplot[
legend cell align={left},
legend style={fill opacity=0.8, draw opacity=1, text opacity=1, draw=lightgray204},
tick align=outside,
tick pos=left,
x grid style={darkgray176},
xlabel={k},
xmin=2.65, xmax=10.35,
xtick style={color=black},
y grid style={darkgray176},
ylabel={Partitions},
ymin=0, ymax=19.9,
ytick style={color=black}
]
\addplot [semithick, forestgreen4416044, opacity=0.7, mark=x, mark size=3, mark options={solid}]
table {%
3 19
4 16
5 9
6 7
7 5
8 3
9 2
10 1
};
\addlegendentry{QuickPartitioner \cite{BQSKitUpdated}}
\addplot [semithick, crimson2143940, opacity=0.7, mark=+, mark size=3, mark options={solid}]
table {%
3 15
4 9
5 6
6 5
7 3
8 3
9 2
10 1
};
\addlegendentry{ScanPartitioner \cite{BQSKitUpdated}}
\addplot [semithick, darkturquoise23190207, opacity=0.7, mark=*, mark size=3, mark options={solid}]
table {%
3 15
4 9
5 6
6 5
7 3
8 3
9 2
10 1
};
\addlegendentry{Clark et al. \cite{PreviousPaper}}
\addplot [semithick, mediumpurple148103189, opacity=0.7]
table {%
3 15
4 9
5 6
6 5
7 3
8 3
9 2
10 1
};
\addlegendentry{Proposed}
\end{groupplot}

\end{tikzpicture}
\end{subfigure}

\vspace{7.5pt}
\hrule
\vspace{7.5pt}

\begin{subfigure}[t]{0.49\textwidth}
	\centering
	\pgfplotsset{every axis legend/.append style={font=\fontsize{9}{8.25}\selectfont}}
	\def\axisdefaultwidth{0.8\textwidth}
	\def\axisdefaultheight{0.55\textwidth}
\begin{tikzpicture}

\definecolor{crimson2143940}{RGB}{214,39,40}
\definecolor{darkgray176}{RGB}{176,176,176}
\definecolor{darkturquoise23190207}{RGB}{23,190,207}
\definecolor{forestgreen4416044}{RGB}{44,160,44}
\definecolor{lightgray204}{RGB}{204,204,204}
\definecolor{mediumpurple148103189}{RGB}{148,103,189}

\begin{groupplot}[group style={group size=1 by 2, x descriptions at=edge bottom, vertical sep = 0pt}]
\nextgroupplot[
log basis y={10},
tick align=outside,
tick pos=left,
title={qft\_20},
unbounded coords=jump,
x grid style={darkgray176},
xmin=2.35, xmax=16.65,
xtick style={color=black},
y grid style={darkgray176},
ylabel={Time (ms)},
ymin=75.0488255335897, ymax=119505296.403975,
ymode=log,
ytick style={color=black},
ytick={0.1,10,1000,100000,10000000,1000000000,100000000000},
yticklabels={
  \(\displaystyle {10^{-1}}\),
  \(\displaystyle {10^{1}}\),
  \(\displaystyle {10^{3}}\),
  \(\displaystyle {10^{5}}\),
  \(\displaystyle {10^{7}}\),
  \(\displaystyle {10^{9}}\),
  \(\displaystyle {10^{11}}\)
}
]
\addplot [semithick, forestgreen4416044, mark=x, mark size=3, mark options={solid}]
table {%
3 235.5513
4 209.29459
5 206.48584
6 219.55733
7 206.81792
8 183.12341
9 206.45003
10 194.05941
11 161.13175
12 159.94289
13 175.48943
14 189.31525
15 183.2886
16 195.30728
};
\addplot [semithick, crimson2143940, mark=+, mark size=3, mark options={solid}]
table {%
3 4613.3793
4 30819.7466
5 152288.3303
6 897227.9409
7 6583728.014
8 62441898.13
9 nan
10 nan
11 nan
12 nan
13 nan
14 nan
15 nan
16 nan
};
\addplot [semithick, darkturquoise23190207, mark=*, mark size=3, mark options={solid}]
table {%
3 304.34398
4 230.84201
5 238.96995
6 274.66821
7 252.84997
8 258.07213
9 272.37
10 262.14255
11 284.31607
12 280.21018
13 267.54065
14 254.21249
15 265.62568
16 260.15696
};
\addplot [semithick, mediumpurple148103189]
table {%
3 194.58123
4 164.09354
5 171.39337
6 161.42089
7 160.514
8 157.11486
9 160.6698
10 154.0997
11 144.22448
12 150.82686
13 163.08544
14 143.63324
15 157.20112
16 151.12785
};

\nextgroupplot[
legend cell align={left},
legend style={fill opacity=0.8, draw opacity=1, text opacity=1, draw=lightgray204},
tick align=outside,
tick pos=left,
unbounded coords=jump,
x grid style={darkgray176},
xlabel={k},
xmin=2.35, xmax=16.65,
xtick style={color=black},
y grid style={darkgray176},
ylabel={Partitions},
ymin=0, ymax=109.05,
ytick style={color=black}
]
\addplot [semithick, forestgreen4416044, opacity=0.7, mark=x, mark size=3, mark options={solid}]
table {%
3 104
4 71
5 51
6 40
7 27
8 18
9 17
10 14
11 8
12 5
13 4
14 3
15 3
16 3
};
\addlegendentry{QuickPartitioner \cite{BQSKitUpdated}}
\addplot [semithick, crimson2143940, opacity=0.7, mark=+, mark size=3, mark options={solid}]
table {%
3 91
4 45
5 33
6 20
7 17
8 10
9 nan
10 nan
11 nan
12 nan
13 nan
14 nan
15 nan
16 nan
};
\addlegendentry{ScanPartitioner \cite{BQSKitUpdated}}
\addplot [semithick, darkturquoise23190207, opacity=0.7, mark=*, mark size=3, mark options={solid}]
table {%
3 92
4 45
5 31
6 20
7 17
8 10
9 9
10 6
11 6
12 5
13 5
14 3
15 3
16 3
};
\addlegendentry{Clark et al. \cite{PreviousPaper}}
\addplot [semithick, mediumpurple148103189, opacity=0.7]
table {%
3 91
4 45
5 35
6 20
7 16
8 10
9 9
10 6
11 6
12 5
13 5
14 3
15 3
16 3
};
\addlegendentry{Proposed}
\end{groupplot}

\end{tikzpicture}
\end{subfigure}
\vrule
\begin{subfigure}[t]{0.49\textwidth}
	\centering
	\pgfplotsset{every axis legend/.append style={font=\fontsize{9}{8.25}\selectfont}}
	\def\axisdefaultwidth{0.8\textwidth}
	\def\axisdefaultheight{0.55\textwidth}
\begin{tikzpicture}

\definecolor{crimson2143940}{RGB}{214,39,40}
\definecolor{darkgray176}{RGB}{176,176,176}
\definecolor{darkturquoise23190207}{RGB}{23,190,207}
\definecolor{forestgreen4416044}{RGB}{44,160,44}
\definecolor{lightgray204}{RGB}{204,204,204}
\definecolor{mediumpurple148103189}{RGB}{148,103,189}

\begin{groupplot}[group style={group size=1 by 2, x descriptions at=edge bottom, vertical sep = 0pt}]
\nextgroupplot[
log basis y={10},
tick align=outside,
tick pos=left,
title={TFIM\_32},
x grid style={darkgray176},
xmin=2.35, xmax=16.65,
xtick style={color=black},
y grid style={darkgray176},
ylabel={Time (ms)},
ymin=682.871556960988, ymax=22901.7408222774,
ymode=log,
ytick style={color=black},
ytick={10,2,5,100,20,50,1000,200,500,10000,2000,5000,100000,20000,50000,1000000,200000,500000},
yticklabels={
  1\(\displaystyle \cdot10^{1}\),
  2\(\displaystyle \cdot10^{0}\),
  5\(\displaystyle \cdot10^{0}\),
  1\(\displaystyle \cdot10^{2}\),
  2\(\displaystyle \cdot10^{1}\),
  5\(\displaystyle \cdot10^{1}\),
  1\(\displaystyle \cdot10^{3}\),
  2\(\displaystyle \cdot10^{2}\),
  5\(\displaystyle \cdot10^{2}\),
  1\(\displaystyle \cdot10^{4}\),
  2\(\displaystyle \cdot10^{3}\),
  5\(\displaystyle \cdot10^{3}\),
  1\(\displaystyle \cdot10^{5}\),
  2\(\displaystyle \cdot10^{4}\),
  5\(\displaystyle \cdot10^{4}\),
  1\(\displaystyle \cdot10^{6}\),
  2\(\displaystyle \cdot10^{5}\),
  5\(\displaystyle \cdot10^{5}\)
}
]
\addplot [semithick, forestgreen4416044, mark=x, mark size=3, mark options={solid}]
table {%
3 1031.15426
4 963.34795
5 999.70433
6 964.03569
7 958.90484
8 944.07263
9 1040.98763
10 1002.25436
11 1002.83704
12 1008.36311
13 1040.24906
14 1018.60336
15 1030.43065
16 983.70206
};
\addplot [semithick, crimson2143940, mark=+, mark size=3, mark options={solid}]
table {%
3 929.42516
4 898.55703
5 920.63926
6 923.43821
7 1014.10021
8 1010.95424
9 1153.06776
10 1288.25887
11 1587.88281
12 2166.96596
13 3421.36117
14 5692.81914
15 10313.69723
16 19522.08713
};
\addplot [semithick, darkturquoise23190207, mark=*, mark size=3, mark options={solid}]
table {%
3 1019.20969
4 1020.41
5 1038.29832
6 1110.64029
7 1179.70345
8 1255.05818
9 1371.1493
10 1464.85758
11 1543.1137
12 1700.02816
13 1857.86198
14 2427.97408
15 2801.33603
16 3401.29035
};
\addplot [semithick, mediumpurple148103189]
table {%
3 873.2425
4 818.71574
5 801.08993
6 806.01779
7 814.77357
8 810.34323
9 818.84305
10 830.58599
11 836.04582
12 927.48193
13 956.21925
14 917.22679
15 849.87391
16 841.17829
};

\nextgroupplot[
legend cell align={left},
legend style={fill opacity=0.8, draw opacity=1, text opacity=1, draw=lightgray204},
tick align=outside,
tick pos=left,
x grid style={darkgray176},
xlabel={k},
xmin=2.35, xmax=16.65,
xtick style={color=black},
y grid style={darkgray176},
ylabel={Partitions},
ymin=0, ymax=262.1,
ytick style={color=black}
]
\addplot [semithick, forestgreen4416044, opacity=0.7, mark=x, mark size=3, mark options={solid}]
table {%
3 250
4 184
5 134
6 108
7 86
8 74
9 66
10 55
11 47
12 41
13 40
14 36
15 27
16 12
};
\addlegendentry{QuickPartitioner \cite{BQSKitUpdated}}
\addplot [semithick, crimson2143940, opacity=0.7, mark=+, mark size=3, mark options={solid}]
table {%
3 241
4 125
5 86
6 57
7 44
8 32
9 28
10 20
11 18
12 14
13 13
14 11
15 10
16 8
};
\addlegendentry{ScanPartitioner \cite{BQSKitUpdated}}
\addplot [semithick, darkturquoise23190207, opacity=0.7, mark=*, mark size=3, mark options={solid}]
table {%
3 241
4 124
5 85
6 58
7 46
8 32
9 28
10 20
11 18
12 14
13 14
14 11
15 10
16 8
};
\addlegendentry{Clark et al. \cite{PreviousPaper}}
\addplot [semithick, mediumpurple148103189, opacity=0.7]
table {%
3 241
4 124
5 84
6 57
7 44
8 32
9 27
10 20
11 18
12 14
13 13
14 11
15 10
16 8
};
\addlegendentry{Proposed}
\end{groupplot}

\end{tikzpicture}
\end{subfigure}
\caption{Performance for Partitioning Methods Across Select Benchmarks}
\label{detailed_performance}
\end{figure*}

\section{Conclusion}
\label{conc}

Quantum circuit optimization presents a viable method for circumventing some of the complexity associated with quantum computing. A promising application of this technique is peephole optimization of quantum circuits, which limits the exponential scaling of synthesis-based optimization. The partitioning algorithm chosen for this application affects both the execution time and quality of the resulting optimized circuit. GTQCP, our proposed method, was compared against three existing partitioning methods designed for this application in a benchmark test and shown to have improved performance against all three. The results show a run time improvement ratio for GTQCP of 18\% against a fast method and 96\% against an exhaustive method. GTQCP also shows a runtime improvement ratio of 70\% against our prior work, which is both fast and high-quality. GTQCP, the exhaustive method, and our prior work all produce nearly identical quality results, with an improvement ratio of 38\% against the fast method. Although GTQCP does not find all possible groups of interacting qubits as our prior work does, the result quality it produces is similar to or better than our prior work in almost all cases, which validates the utility of the new heuristic. The proposed method also addresses a limitation of our prior work which impairs performance on shallow circuits.

Future work on GTQCP should seek to tighten the upper bound on the time complexity of the algorithm, which would also likely yield a better understanding of the limitations, if any, of this algorithms compared to exhaustive approaches. GTQCP would also benefit from improvements to overall maturity of the method, such as adding support for more advanced scoring capabilities like a lookahead mechanism. Similarly, integration with synthesis tools would enable the algorithm to select partitions for performance directly, rather than just optimizing for partition size.

\section*{References}

\printbibliography[heading=none]

\end{document}